\documentclass{article}

\usepackage{amsmath}

\newtheorem{claim}{Claim}
\newtheorem{prop}{Proposition}
\newtheorem{counter}{Counterexample}

\author{Blake Hegerle}
\title{A Counterexample to a Proposed Proof of P=NP by S. Gubin}

\begin{document}
\maketitle

In \cite{gubin}, the claim is put forth that P=NP; the form of this
claim is an algorithm which purportedly can solve the 3SAT problem in
$\mathrm O(n^4)$ time.

The 3SAT problem (or ``3-SAT problem,'' as it is refered to in
\cite{gubin}) is to determine if the formula
\begin{equation}
  \label{eq:formula}
  d_1 \wedge d_2 \wedge \dotsb \wedge d_m
\end{equation}
is satisfiable, where each clause $d_k$ with $1 \le k \le m$ is a
disjunction of at most three variables or their negations from the set
\begin{equation}
  \label{eq:variable set}
  B=\{b_1, b_2,\dotsc, b_n\}.
\end{equation}

The validity of the algorithm rests on the
following claim:
\begin{claim}
  Let \eqref{eq:formula} and \eqref{eq:variable set} be the given
  instance of 3SAT. Let $C$ be the set of clauses of the instance:
  \[
  C = \{d_{1}, d_{2}, \dotsc, d_{m} \}.
  \]
  The instance is non-satisfiable if and only if at least one of the
  following is true:
  \begin{description}
  \item[Pattern 1.]  There is $\alpha \in B$:
    \[
    \{\alpha,~ \bar{\alpha}\} \subseteq C;
    \]
  \item[Pattern 2.]
    There are different $\alpha, \beta \in B$:
    \[
    \{\alpha \vee \beta,~ \alpha \vee \bar{\beta},~ \bar{\alpha} \vee
    \beta,~ \bar{\alpha} \vee \bar{\beta} \} \subseteq C;
    \]
  \item[Pattern 3.]
    There are different $\alpha, \beta, \gamma \in B$:
    \[
    \begin{array}{l}
      \{\alpha \vee \beta \vee \gamma,~ \alpha \vee \beta \vee
      \bar{\gamma},~ \alpha \vee \bar{\beta} \vee \gamma,~ \alpha \vee
      \bar{\beta} \vee \bar{\gamma}, \\ ~\bar{\alpha} \vee \beta \vee
      \gamma,~ \bar{\alpha} \vee \beta \vee \bar{\gamma},~
      \bar{\alpha} \vee \bar{\beta} \vee \gamma,~ \bar{\alpha} \vee
      \bar{\beta} \vee \bar{\gamma} \} \subseteq C.
    \end{array}
    \]
  \end{description}
\end{claim}

This claim is incorrect. The proof supplied in \cite{gubin} only
addresses the ``if'' direction; that is, the following Proposition is
proved, which \emph{is} true.

\begin{prop}
\label{claim:sufficient for 3SAT}
Let \eqref{eq:formula} and \eqref{eq:variable set} be the given
  instance of 3SAT. Let $C$ be the set of clauses of the instance:
  \[
  C = \{d_{1}, d_{2}, \dotsc, d_{m} \}.
  \]
  The instance is non-satisfiable if any of the following are true:
  \begin{enumerate}
  \item 
\label{item:alpha and not alpha}
 There is $\alpha \in B$:
    \[
    \{\alpha,~ \bar{\alpha}\} \subseteq C;
    \]
  \item
\label{item:alpha beta combinations}
    There are different $\alpha, \beta \in B$:
    \[
    \{\alpha \vee \beta,~ \alpha \vee \bar{\beta},~ \bar{\alpha} \vee
    \beta,~ \bar{\alpha} \vee \bar{\beta} \} \subseteq C;
    \]
  \item
\label{item:alpha beta gamma combinations}
    There are different $\alpha, \beta, \gamma \in B$:
    \[
    \begin{array}{l}
      \{\alpha \vee \beta \vee \gamma,~ \alpha \vee \beta \vee
      \bar{\gamma},~ \alpha \vee \bar{\beta} \vee \gamma,~ \alpha \vee
      \bar{\beta} \vee \bar{\gamma}, \\ ~\bar{\alpha} \vee \beta \vee
      \gamma,~ \bar{\alpha} \vee \beta \vee \bar{\gamma},~
      \bar{\alpha} \vee \bar{\beta} \vee \gamma,~ \bar{\alpha} \vee
      \bar{\beta} \vee \bar{\gamma} \} \subseteq C.
    \end{array}
    \]
  \end{enumerate}
\end{prop}

The supposition is a sufficient but not necessary condition for a
given formula to lack a solution. It is easy to find a counterexample.

\begin{counter}
The formula
\[
\bar a \wedge \bar b \wedge \bar c \wedge (a \vee b) \wedge (a \vee c)
\wedge (b \vee c)
\]
is not satisfiable, even though it does not meet either condition
\ref{item:alpha and not alpha}, \ref{item:alpha beta
combinations}, or \ref{item:alpha beta gamma combinations} of
Claim~\ref{claim:sufficient for 3SAT}.
\end{counter}

\bibliography{counterexample} 
\bibliographystyle{plain}

\end{document}